\documentclass[aps,preprint,nofootinbib,floatfix]{revtex4}
\usepackage{graphicx}
\usepackage{amsmath,amssymb}
\usepackage{url}
\usepackage{epstopdf}
\newcommand{\lsim}{\mathrel{\mathop{\kern 0pt \rlap
  {\raise.2ex\hbox{$<$}}}
  \lower.9ex\hbox{\kern-.190em $\sim$}}}
\newcommand{\gsim}{\mathrel{\mathop{\kern 0pt \rlap
  {\raise.2ex\hbox{$>$}}}
  \lower.9ex\hbox{\kern-.190em $\sim$}}}

\begin{document}
\title{Cosmic Antiproton Constraints on Effective Interactions\\
 of the Dark Matter}
\author{Kingman Cheung$^{1,2,3}$, Po-Yan Tseng$^{2}$, and Tzu-Chiang Yuan$^4$}

\affiliation{
$^1$Division of Quantum Phases \& Devices, School of Physics, 
Konkuk university, Seoul 143-701, Korea \\
$^2$Department of Physics, National Tsing Hua University, 
Hsinchu 300, Taiwan
\\
$^3$Physics Division, National Center for Theoretical Sciences,
Hsinchu 300, Taiwan
\\
$^4$Institute of Physics, Academia Sinica, Nankang, Taipei 11529, Taiwan
}

\date{\today}

\begin{abstract}
Using an effective interaction approach to describe the 
interactions between the dark matter particle and the light degrees 
of freedom of the standard model,
we calculate the antiproton flux due to the annihilation of the
dark matter in the Galactic Halo and compare to the most recent 
antiproton spectrum of the PAMELA experiment. We obtain useful constraints on the size
of the effective interactions that are comparable to those deduced from collider
and gamma-ray experiments.
\end{abstract}
\maketitle

\section{Introduction}
The presence of cold dark matter (CDM) in our Universe is now well established
by a number of observational experiments, especially the very precise 
measurement of the cosmic microwave background radiation
in the Wilkinson Microwave Anisotropy Probe (WMAP) experiment \cite{wmap}.
The measured value of the CDM relic density is
\[
 \Omega_{\rm CDM}\, h^2 = 0.1099 \;\pm 0.0062 \;,
\]
where $h$ is the Hubble constant in units of $100$ km/Mpc/s.
Though the gravitation nature of the dark matter is established, 
we know almost nothing about its particle nature, except that it is
nonbaryonic and to a high extent electrically neutral.

One of the most appealing and natural CDM particle candidates is 
{\it weakly-interacting massive particle} (WIMP).  It is a coincidence
that if the dark matter (DM) is thermally produced  in the early Universe,
the required annihilation cross section is right at the order of
weak interaction.  The relation between the relic density and
the thermal annihilation cross section can be given by the following 
simple formula \cite{hooper}
\begin{equation}
\label{rate}
\Omega_\chi h^2 \simeq \frac{ 0.1 \;{\rm pb} }{\langle \sigma v \rangle} \;,
\end{equation}
where $\langle \sigma v \rangle$ is the 
annihilation rate of the dark matter around the time of freeze-out.
Given the measured $\Omega_{\rm CDM} h^2$ the annihilation
rate is about $1$ pb or $10^{-26}\;{\rm cm}^3 \, {\rm s}^{-1}$.  
This is exactly the size of the cross section that one expects from a weak interaction
process and that would also give a large to moderate production rate of the WIMP at the 
Large Hadron Collider (LHC).
In general, production of dark matter at the LHC would give rise to
a large missing energy.  Thus, the anticipated signature in the 
final state is high-$p_T$ jets or leptons plus a large missing energy.
Note that there could be nonthermal sources for the dark matter, such 
as decay from exotic relics like moduli fields, cosmic strings, etc. In 
such cases, the annihilation rate in Eq.~(\ref{rate}) can be larger than
the value quoted above.

There have been many proposed candidates for the dark matter. 
Instead of specifying a particular model we adopt an effective interaction
approach to describe the interactions of the dark matter particle with
the standard model (SM) particles \cite{cao,bai,tait,tait-gamma,fan,ours}.
One simple realization is that the dark matter particle exists in a 
hidden sector, which communicates to the SM sector via a heavy degree
of freedom in the connector sector. At energy scale well below this heavy mediator the interactions
can be conveniently described by a set of effective interactions.
The strength of each interaction depends on  the nature of the dark
matter particle and the mediator.
An interesting set of interactions between the fermionic dark matter $\chi$ 
and the light quarks $q$
can be described by 
$(\bar q \Gamma q)(\bar \chi \Gamma^\prime \chi)$, where
$\Gamma , \Gamma^\prime = \sigma^{\mu\nu},\,\sigma^{\mu\nu}\gamma^5,\, \gamma^\mu,\, 
\gamma^\mu\gamma^5,\,\gamma^5$ and 1. 
Note that due to the following identity
$$
\sigma^{\mu\nu} \gamma^5 = \frac{i}{2} \epsilon^{\mu\nu\alpha\beta} \sigma_{\alpha\beta} \; ,
$$
the axial tensor $\sigma^{\mu\nu} \gamma^5$ is related to 
the tensor $\sigma^{\alpha\beta}$ and thus should not be regarded as 
an independent set.  
A more complete set of interactions involving
fermionic and scalar dark matter candidates
that we will study in this work are listed in Table \ref{table1}. 
Without a particular model in mind we will treat each interaction
independently in our analysis.

There have been some recent works on constraining the interactions 
at present and
future collider experiments \cite{cao,bai,tait} and using gamma-ray experiments
\cite{tait-gamma}. 
Fan {\it et al.} \cite{fan}
also wrote down the effective nonrelativistic interactions between the dark
matter and nuclei.
There was another work in which the dark matter couples
only to the top quark and corresponding predictions at direct and indirect
detection experiments as well as colliders are obtained \cite{ours}.

In the present work, we first estimate the lower bounds on the 
new interactions based on the fact that if a particular interaction
is the only contribution that can thermalize the 
dark matter particle in the early Universe, we require this 
interaction must be strong enough such that the resulting relic density 
would not overclose the Universe.  
In addition, we proceed to calculate the antiproton flux coming from 
the effective interactions.  We expect the latest antiproton data 
from PAMELA \cite{pamela-p}
can put a strong constraint on the size of the interactions, 
based on the fact that the existing data do not allow excessive flux 
above the conventional background. On the other hand, the positron 
spectrum from PAMELA \cite{pamela-e} showed some excessive above the 
conventional background and thus if we used it to constrain the model,
it in general gives a weaker constraint than the antiproton flux
\cite{gauge-higgs,ours}.  Therefore, in this work we focus on the
constraints from antiproton flux.
Similar ideas of using the antiproton flux from earlier PAMELA data 
was considered in Ref. \cite{keung} to confront the low energy CoGeNT 
experiment \cite{cogent} for a lower mass DM.

The organization of the paper is as follows.  In the next section, we describe 
the interactions
 between the dark matter particle and the SM particles, in particular 
quarks and gluons.
In Sec. III, we study the velocity dependence of the effective operators based on 
the nonrelativistic reduction.
In Sec. IV, we calculate the annihilation rates during the freeze-out
and make sure that the interactions would not overclose the Universe.
In Sec. V, we calculate the antiproton spectrum due to the
dark matter annihilation in Galactic Halo. 
We compare with other constraints and conclude in Sec. VI.

\section{Effective Interactions}

Let us start by assuming the dark matter is a Dirac fermion and
its effective interactions with light quarks via a (axial) vector-boson 
or tensor-type 
exchange are given by the following dimension 6 operators
\begin{equation}
\label{eff-q}
{\cal L}_{i=1-6} = \frac{C}{\Lambda_i^2} \,
   \left ( \overline{\chi} \Gamma_1 \chi \right )\;
           \left ( \bar{q} \Gamma_2 q \right ) \;,
\end{equation}
where  
$\Gamma_{1,2} = \gamma^\mu , \gamma^\mu \gamma^5, \sigma^{\mu\nu}$ or 
$\sigma^{\mu\nu} \gamma^5$
with 
$\sigma^{\mu\nu} \equiv i (\gamma^\mu \gamma^\nu -\gamma^\nu \gamma^\mu ) /2$, and
$C$ is an effective coupling constant of order $O(1)$.
For Majorana fermion the $\Gamma_1 = \gamma^\mu$ or $\sigma^{\mu\nu}$ type 
interaction is identically zero, and so for vector or tensor type interaction 
the fermion $\chi$ in Eq.(\ref{eff-q}) is understood to be Dirac. 
Explicitly, we 
assume the dark matter candidate to be Dirac, but the results
are also applicable to Majorana dark matter.  

Next set of operators are associated with (pseudo) scalar-boson-type exchange
\begin{equation}
{\cal L}_{i=7-10} = \frac{C m_q }{\Lambda_i^3} \,
   \left ( \overline{\chi} \Gamma_1 \chi \right )\;
           \left ( \bar{q} \Gamma_2 q \right ) \;,
\end{equation}
where $\Gamma_{1,2} = 1$ or $i\gamma^5$.  The $m_q$ dependence in the
coupling strength is explicitly shown for scalar-type interactions.
We use the current quark masses in the Lagrangian given by \cite{pdg}:
\begin{eqnarray}
 m_u & = & 0.0025\; {\rm GeV}, \quad  m_d \; = \; 0.005 \;{\rm GeV}, \quad
 m_s \; = \; 0.101 \;{\rm GeV}, \nonumber \\
 m_c & = &1.27 \;{\rm GeV}, \quad
 m_b \; = \; 4.19 \; {\rm GeV}, \quad 
 m_t \; = \; 172 \; {\rm GeV}.
\nonumber
\end{eqnarray}
Another light degree of freedom that couples to the Dirac dark matter 
is the gluon field 
\footnote
{We do not study the other gauge bosons, like $W$ and $Z$ bosons, because
they decay into light quarks which then fragment into $\bar p$.  
The secondary antiproton spectrum would be softer in this case.}
\begin{eqnarray}
{\cal L}_{i=11-12} &=& \frac{C \alpha_s(2 m_\chi) }{ 4 \Lambda_i^3} \,
   \left ( \overline{\chi} \Gamma \chi \right )\;
          G^{a\mu\nu} G^a_{\mu \nu} \\
{\cal L}_{i=13-14} &=& \frac{C \alpha_s(2 m_\chi) }{4 \Lambda_i^3} \,
   \left ( \overline{\chi} \Gamma \chi \right )\;
          G^{a\mu\nu} \tilde{G}^a_{\mu \nu} 
\end{eqnarray}
where $\Gamma = 1$ or $i\gamma^5$ and the strong coupling constant 
is evaluated at the scale $2m_\chi$ 
where $m_\chi$ is the dark matter mass.

Finally, we also write down the corresponding operators for complex
scalar dark matter.  Again, we note that the interactions for real scalar
dark matter is similar to complex one and differ by a factor of two. 
We  simply focus on the complex scalar dark matter.  The operators 
corresponding to vector boson exchange are
\begin{equation}
{\cal L}_{i=15,16} =  \frac{C}{\Lambda_i^2} \,
   \left ( \chi^\dagger \overleftrightarrow{\partial_\mu} \chi \right )\;
           \left ( \bar{q} \gamma^\mu \Gamma q \right ) \;,
\end{equation}
where $\Gamma = 1$ or $\gamma^5$ and 
$\chi^\dagger \overleftrightarrow{\partial_\mu} \chi =
 \chi^\dagger (\partial_\mu \chi) - (\partial_\mu \chi^\dagger) \chi$.  
Those corresponding to a scalar boson exchange are
\begin{equation}
{\cal L}_{i=17,18} =  \frac{C m_q }{\Lambda_i^2} \,
   \left (  \chi^\dagger  \chi \right )\;
           \left ( \bar{q} \Gamma q \right ) \;,
\end{equation}
where $\Gamma = 1$ or $i\gamma^5$.  The corresponding gluonic operators are
\begin{eqnarray}
{\cal L}_{i=19} &=& \frac{C \alpha_s(2 m_\chi) }{4 \Lambda_i^3} \,
   \left ( \chi^\dagger \chi \right )\;
        G^{a \mu\nu} G_{a \mu \nu} \; , \\
{\cal L}_{i=20} &=& \frac{i C \alpha_s(2 m_\chi) }{4 \Lambda_i^3} \,
   \left ( \chi^\dagger \chi \right )\;
        G^{a\mu\nu} \tilde{G}_{a\mu \nu} \; . 
\end{eqnarray}
The whole list of operators are listed in Table~\ref{table1}.
We will consider one operator at a time, 
and set the coefficient $C=1$ for simplicity.

Note that in calculating the annihilation rate in the freeze-out in 
the early Universe, we include all light-quark flavors ($u,d,s,c,b$) 
as well as the heavy top quark which is relevant when 
$m_\chi$ rises above the top quark threshold.  
However, in the calculation of the antiproton flux from dark matter 
annihilation in the present Galactic halo, we only include the light-quark
flavors.  We ignore the 
$\chi \overline{\chi} \to t \bar t$ contribution, 
because the $t$ and $\bar t$ first decay into $b W \to b
q \bar q'$ before each light quark undergoes fragmentation into 
hadrons, including proton and antiproton.
Therefore, the antiproton spectrum would be significantly softer than the
direct fragmentation as in $\chi \overline{\chi} \to q \bar q$ \cite{ours}. 
We anticipate that by ignoring the $t \bar t$ contribution
the limits we obtain from the PAMELA data would not be
affected to any significant amount in the case we just use 
the five light-quark flavors.

\begin{table}[th!]
\caption{\small \label{table1}
The list of effective interactions between the dark matter and the light
degrees of freedom (quark or gluon). 
We have suppressed the color index
on the quark and gluon fields. These operators have also been
analyzed in Refs.~\cite{cao,tait,tait-gamma}.  }
\medskip
\begin{tabular}{lr}
\hline
Operator & Coefficient \\
\hline
\hline
\multicolumn{2}{c}{Dirac DM, Vector Boson Exchange} \\
\hline
$O_1 = (\overline{\chi} \gamma^\mu \chi)\, (\bar q \gamma_\mu q)$ &
                    $ \frac{C}{\Lambda^2}$ \\
$O_2 = (\overline{\chi} \gamma^\mu \gamma^5\chi)\, (\bar q \gamma_\mu  q)$ &
                    $ \frac{C}{\Lambda^2}$ \\
$O_3 = (\overline{\chi} \gamma^\mu \chi)\, (\bar q \gamma_\mu \gamma^5 q)$ &
                     $\frac{C}{\Lambda^2}$ \\
$O_4 = (\overline{\chi} \gamma^\mu \gamma^5 \chi)\, 
    (\bar q \gamma_\mu \gamma^5 q)$ &  $   \frac{C}{\Lambda^2}$ \\
$O_5 = (\overline{\chi} \sigma^{\mu\nu} \chi)\, (\bar q \sigma_{\mu\nu} q)$ & 
                     $\frac{C}{\Lambda^2} $\\
$O_6 = (\overline{\chi} \sigma^{\mu\nu} \gamma^5 \chi)\, 
  (\bar q \sigma_{\mu\nu} q)$ &  $\frac{C}{\Lambda^2} $\\
\hline
\multicolumn{2}{c}{Dirac DM, Scalar Boson Exchange} \\
\hline
$O_7 = (\overline{\chi}  \chi)\, (\bar q  q)$ & $ \frac{C m_q }{\Lambda^3}$ \\
$O_8 = (\overline{\chi} \gamma^5  \chi)\, (\bar q  q)$ &  
                          $\frac{i  C m_q }{\Lambda^3}$ \\
$O_9 = (\overline{\chi}  \chi)\, (\bar q \gamma^5 q)$&
              $\frac{i C m_q }{\Lambda^3}$ \\
$O_{10} = (\overline{\chi} \gamma^5 \chi)\, (\bar q \gamma^5 q) $ &  
                        $ \frac{C m_q }{\Lambda^3}$ \\
\hline
\multicolumn{2}{c}{Dirac DM, Gluonic} \\
\hline
$O_{11} = (\overline{\chi} \chi)\, G_{\mu\nu} G^{\mu\nu}$ & 
                  $       \frac{C \alpha_s }{4 \Lambda^3} $ \\
$O_{12} = (\overline{\chi} \gamma^5 \chi)\, G_{\mu\nu} G^{\mu\nu} $ & 
                $         \frac{i C \alpha_s }{4 \Lambda^3} $\\
$O_{13} = (\overline{\chi} \chi)\, G_{\mu\nu} \tilde{G}^{\mu\nu}$ & 
                         $\frac{C \alpha_s }{4 \Lambda^3} $\\
$O_{14} = (\overline{\chi} \gamma^5 \chi)\, G_{\mu\nu} \tilde{G}^{\mu\nu} $& 
                        $ \frac{i C \alpha_s }{4 \Lambda^3} $\\
\hline
\hline
\multicolumn{2}{c}{Complex Scalar DM, Vector Boson Exchange} \\
\hline
$O_{15} = (\chi^\dagger \overleftrightarrow{\partial_\mu} \chi)\, 
        ( \bar q \gamma^\mu q ) $&
                     $    \frac{ C }{ \Lambda^2} $\\
$O_{16} = (\chi^\dagger \overleftrightarrow{\partial_\mu} \chi)\, 
  ( \bar q \gamma^\mu \gamma^5 q ) $&
                      $   \frac{ C }{ \Lambda^2} $\\
\hline
\multicolumn{2}{c}{Complex Scalar DM, Scalar Vector Boson Exchange} \\
\hline
$O_{17} = (\chi^\dagger \chi)\, ( \bar q  q ) $&
                       $  \frac{C m_q }{ \Lambda^2}$ \\
$O_{18} = (\chi^\dagger \chi)\, ( \bar q  \gamma^5 q ) $&
                $         \frac{i C m_q }{ \Lambda^2} $ \\
\hline
\multicolumn{2}{c}{Complex Scalar DM, Gluonic} \\
\hline
$O_{19} = (\chi^\dagger \chi)\, G_{\mu\nu} {G}^{\mu\nu} $   & 
                       $  \frac{C \alpha_s }{4 \Lambda^2} $\\
$O_{20} = (\chi^\dagger \chi)\, G_{\mu\nu} \tilde{G}^{\mu\nu} $& 
                      $   \frac{i C \alpha_s }{4 \Lambda^2} $\\
\hline
\hline
\end{tabular}
\end{table}

\section{Velocity Dependence in the Nonrelativistic limits}

In order to easily understand the results that we obtain in Sec. V, we
are going to examine the dependence of the annihilation cross section 
on the velocity of the dark matter particle in the nonrelativistic limit.
The current velocity of the dark matter in the Universe around the Sun
is about $v\approx 300\, {\rm km}\, {\rm s}^{-1} \approx 10^{-3} c$, 
where $c$ is the speed of the light.  Given such a small $v$ the dependence
on $v$ is very important.  For instance, the annihilation rate of 
the Dirac DM with a scalar boson exchange, given by the interaction 
in $O_7$, would suffer from a factor of $v^2$.  Therefore, we expect the 
antiproton flux from such an operator would be very small.

Let us consider the operators $O_1$ to $O_6$ with 
(axial) vector-boson/tensor-like exchange.
In terms of Dirac spinors ($\psi$ and $\bar \psi$) the relevant part of 
the annihilation amplitude of the Dirac DM is given by 
\begin{equation}
  \bar \psi(p_2) \Gamma \psi(p_1) 
\end{equation}
for $\overline{\chi}(p_2) \chi(p_1) \to q \bar q$ and $\Gamma = \gamma^\mu,
\gamma^\mu \gamma^5, \sigma^{\mu\nu}$ or $\sigma^{\mu\nu} \gamma^5$.
In Dirac representation, the gamma matrices are given by
\[
  \gamma^0 = \left( \begin{array}{cc} 
                  I & 0 \\
                  0 & - I
              \end{array} \right ),\;\;
  \gamma^i = \left( \begin{array}{cc} 
                  0 & \sigma_i \\
                  -\sigma_i & 0
              \end{array} \right ),\;\;
  \gamma^5 = \left( \begin{array}{cc} 
                  0 & I \\
                  I & 0
              \end{array} \right ) \,,
\]
where $\sigma_i (i=1,2,3)$ are the Pauli matrices.  
In the nonrelativistic limit, the spinor for the DM $\chi$ is 
$\psi = \xi \left( \begin{array}{c}
                  1 \\
                  \epsilon 
               \end{array} \right )$, where $\epsilon = O(v/c)$.
On the other hand, the spinor for the antiparticle $\overline{\chi}$ 
is  $\bar \psi = \eta^\dagger ( \epsilon,   1) \gamma^0$. 
\footnote
{It is different from the direct scattering with a nucleon, where
we need the $\bar \psi = \xi^\dagger (1, \epsilon) \gamma^0$.
}
Therefore, we can expand $\bar \psi \gamma^\mu \psi$,
in the nonrelativistic limit, as
\begin{eqnarray}
  \bar \psi \gamma^0 \psi &\simeq& 2 \epsilon \eta^\dagger  \xi \, \nonumber \\
  \bar \psi \gamma^i \psi &\simeq & (1+\epsilon^2)
                         \eta^\dagger \sigma_i \xi \, \nonumber 
\end{eqnarray}
where the space-like parts are not suppressed by $v/c$.  On the other 
hand, $\bar \psi \gamma^\mu \gamma^5\psi$ in the nonrelativistic limit are 
\begin{eqnarray}
  \bar \psi \gamma^0\gamma^5 \psi &\simeq& (1+\epsilon^2)
                               \eta^\dagger  \xi \, \nonumber \\
  \bar \psi \gamma^i \gamma^5 \psi &\simeq& 2 \epsilon \eta^\dagger \sigma_i \xi 
  \, \nonumber 
\end{eqnarray}
where the space-like parts are suppressed by $v/c$. 
It is clear that in the nonrelativistic limit the time-like and space-like
parts behave very differently.  We can then consider them separately 
when it is squared, traced, and contracted with the trace of the light
quark leg. 
If we look at the trace of the part $(\bar q \gamma^\mu q)$ or 
$(\bar q \gamma^\mu \gamma^5 q)$ in the annihilation amplitude, the
time-like part after being squared and traced gives a quantity close to
zero, while the space-like part after squared and traced gives a quantity
in the order of $m_\chi^2$.  Therefore, it is clear now that 
$\bar \psi \gamma^\mu \psi$ multiplied to $(\bar q \gamma_\mu q)$ or 
$(\bar q \gamma_\mu \gamma^5 q)$ will not be suppressed, while
$\bar \psi \gamma^\mu \gamma^5\psi$ multiplied to $(\bar q \gamma_\mu q)$ or 
$(\bar q \gamma_\mu \gamma^5 q)$ will be suppressed. The above observation
is consistent with the results that we obtain in Sec. V.  From 
Table~\ref{limit} the limits on $O_1$ and $O_3$ are much stronger than
those on $O_2$ and $O_4$.  The operators $O_5$ and $O_6$ contain unsuppressed
components in $\mu\nu = 0 i$ entries.

In contrast, the operators $O_7$ to $O_{10}$ with (pseudo) scalar-boson 
exchange 
are suppressed when there is no $\gamma^5$ in the fermion line of $\chi$,
which is obvious from the following in the nonrelativistic limit
\begin{eqnarray}
  \bar \psi  \psi &\sim&  \epsilon \eta^\dagger  \xi \; , \nonumber \\
  \bar \psi  \gamma^5 \psi &\sim&  \eta^\dagger  \xi  \; \; .
  \nonumber 
\end{eqnarray}
Again, it is then obvious from Table~\ref{limit} that the limits on
$O_8$ and $O_{10}$ are much stronger than those on $O_7$ and $O_9$.
The gluonic operators in $O_{11-14}$ follow similar patterns: 
$O_{11,13}$ are suppressed relative to $O_{12,14}$. 

It is also straightforward to understand the velocity dependence for
the scalar DM, represented by the operators $O_{15-20}$.  Except for
$O_{15,16}$ all of them are not suppressed by $v/c$, because of the 
presence of the $\overleftrightarrow{\partial_\mu}$ in $O_{15,16}$.  
This $\partial_\mu$ will bring down $p_\mu$ in the  vertex factor. 
While $p_0$ is of order $m_\chi$, $p_i$ is $v/c$.  Therefore, when it
contracts with the quark leg, the overall result is suppressed by $v/c$.

\section{Annihilation rates around the freeze-out}

It is obvious from Eq.~(\ref{rate}) that if the annihilation rate 
falls below 1 pb, then the thermal relic density would be more than the
WMAP data can allow.  Therefore, we have to restrict the annihilation rate
to be larger than about 1 pb.  More precisely,
using the most recent WMAP result on dark matter density $\Omega_{\rm CDM}
h^2 = 0.1099 \pm 0.0062$ \cite{wmap} the annihilation rate is 
\begin{equation}
\langle \sigma v \rangle \simeq 0.91 \; {\rm pb} \;\; .
\end{equation}
We assume $v \approx 0.3 $ at around the freeze-out time in the early 
Universe.  

\begin{figure}[th!]
\centering
\includegraphics[width=5in]{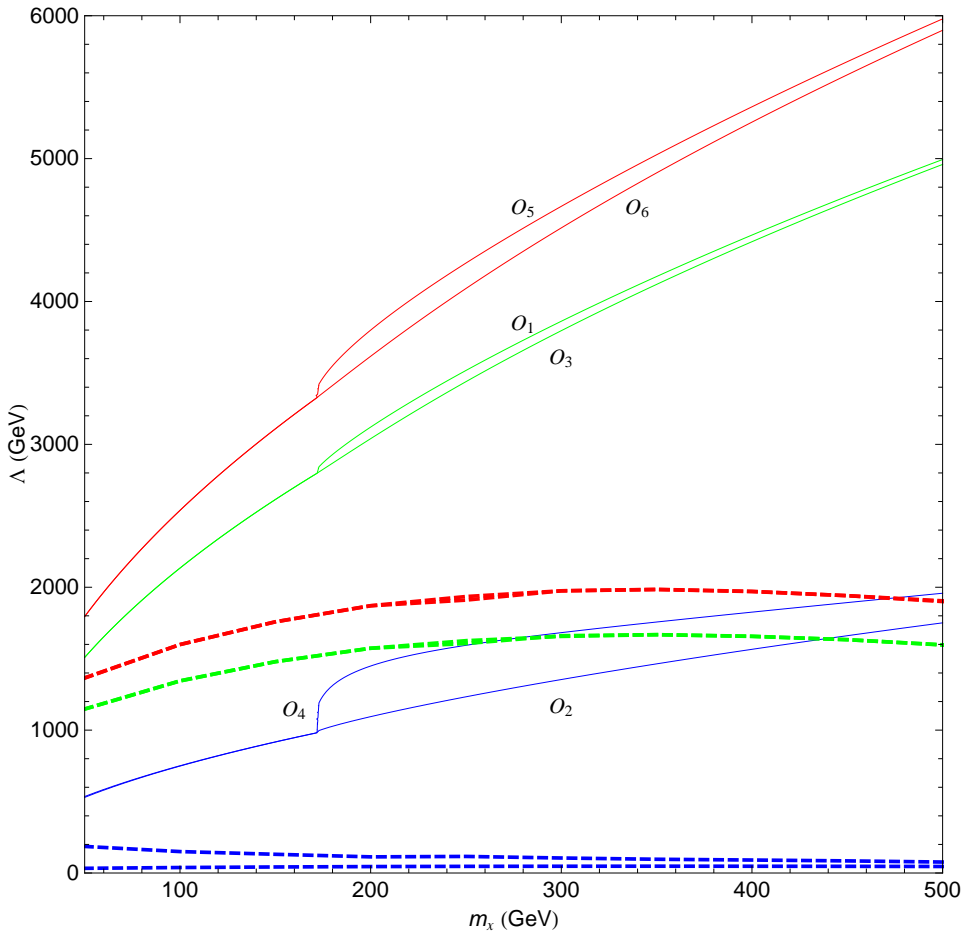}
\caption{\small \label{figA}
The upper limits on $\Lambda_i$ due to the relic density constraint,
requiring $\sigma v \ge 0.91$ pb for operators $O_{1-6}$ involving
Dirac DM with (axial) vector-boson/tensor-like exchanges (shown by solid lines).
The lower limits on $\Lambda_i$ due to the antiproton-flux constraint
at $3\sigma$ level for the same operators (shown by dashed lines with
the corresponding color).
}
\end{figure}

\begin{figure}[th!]
\centering
\includegraphics[width=5in]{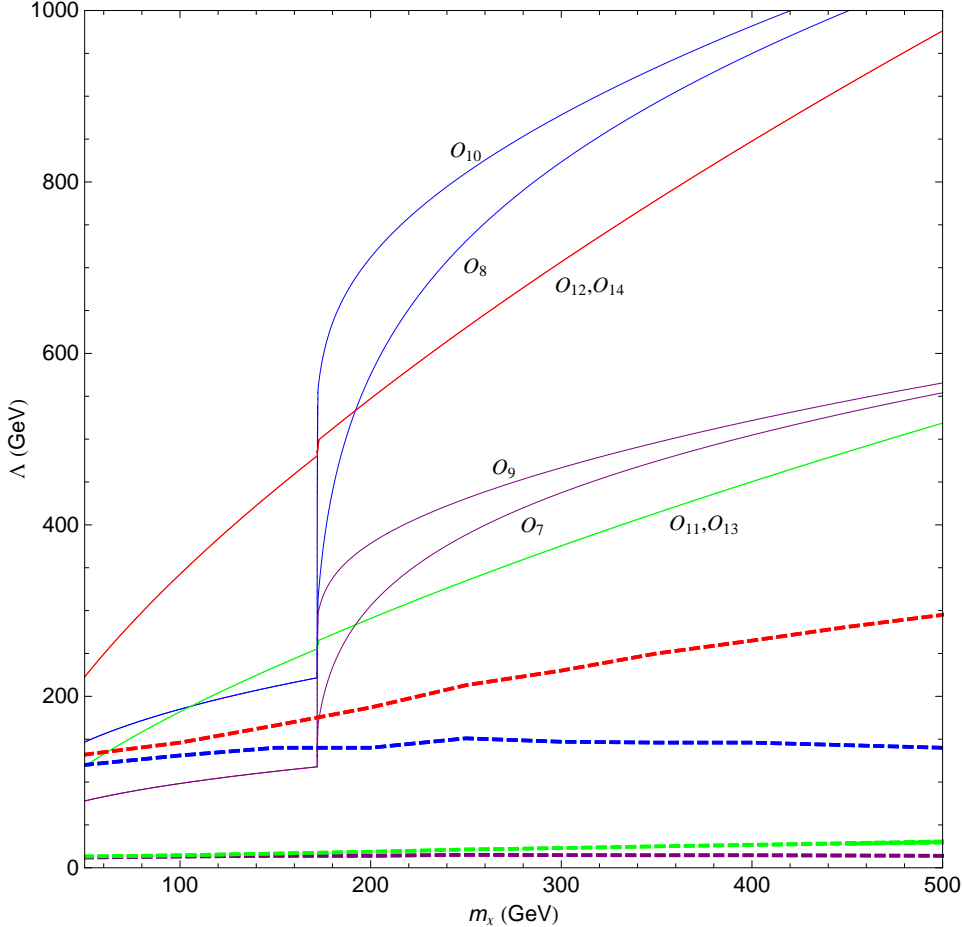}
\caption{\small \label{figB}
The same as Fig.~\ref{figA}, but for operators $O_{7-14}$ involving
Dirac DM with (pseudo) scalar-boson exchanges ($O_{7-10}$) and Dirac DM with
gluonic interactions ($O_{11-14}$).
}
\end{figure}

\begin{figure}[th!]
\centering
\includegraphics[width=5in]{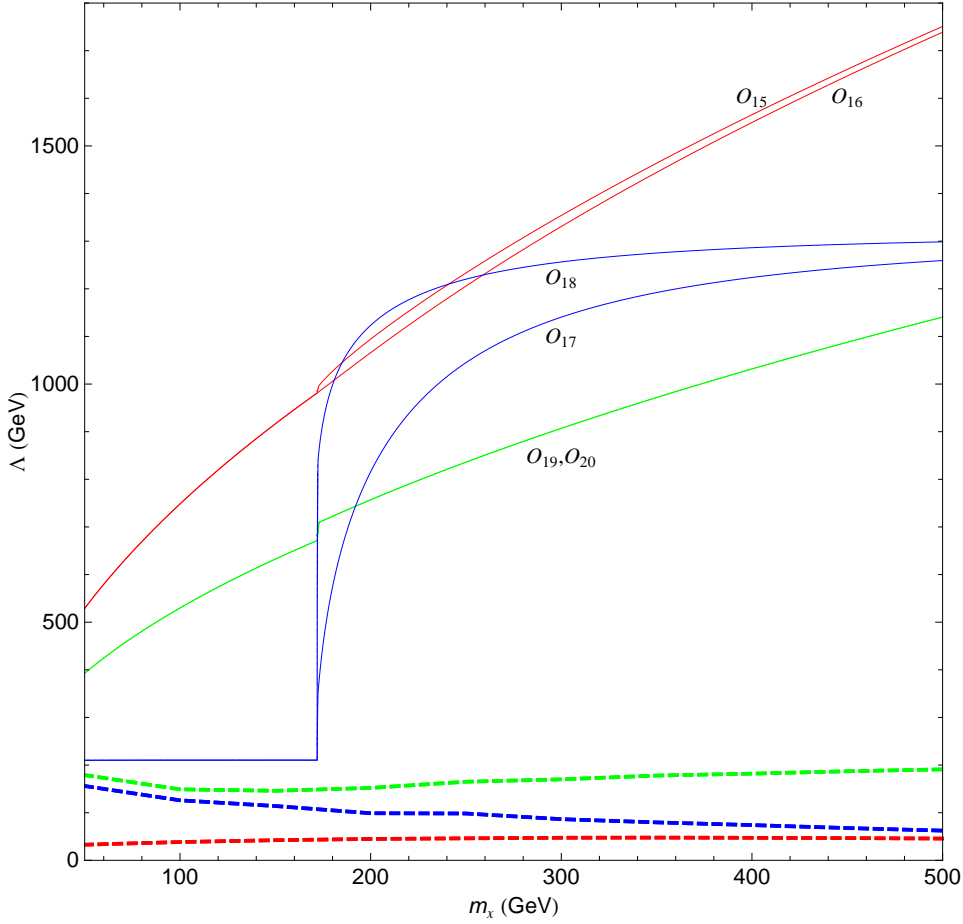}
\caption{\small \label{figC}
The same as Fig.~\ref{figA}, but for operators $O_{15-20}$ involving
complex scalar DM.
}
\end{figure}

We calculate the annihilation rates for all the operators and show the
contours in Fig.~\ref{figA}, Fig.~\ref{figB}, and Fig.~\ref{figC} for 
Dirac DM with (axial) vector-boson/tensor-like exchanges, 
Dirac DM with (pseudo) scalar-boson exchanges and 
Dirac DM with gluonic interactions, and scalar DM,
respectively.  The solid lines are the contours in $(m_\chi, \Lambda)$
plane with annihilation rate $\sigma v = 0.91$ pb.  All the values of
$\Lambda$ above the solid lines would give a too small annihilation rate,
and thus would result in a too large thermal relic density. Therefore, 
the $\Lambda$ below the solid lines is the allowed region. 
The cusp structures in the plots are due to the onset of the top quark
contributions when $m_\chi > m_t$.
The dashed lines are the limits from the antiproton flux, which
will be explained in the next section.

\section{Antiproton Flux}

An important method to detect the dark matter is by measuring 
its annihilation products in Galactic halo. Current experiments can
detect the positron, antiproton, gamma ray, and deuterium from
dark matter annihilation.   The Milky Way halo may contain
clumps of dark matter, from where the annihilation of dark matter particles
may give rise to large enough signals, such as positron and antiproton,
that can be identified by a number of antimatter search experiments. 
The most recent ones come from PAMELA \cite{pamela-e,pamela-p}, 
which showed a spectacular
rise in the positron spectrum but an expected spectrum for antiproton. 
It may be due to nearby pulsars or dark matter annihilation or decays.
If it is really due to dark matter annihilation, the dark matter
would have very strange properties, because it only gives positrons in 
the final products but not antiproton.  Here we adopt a conservative 
approach. We use the observed antiproton as a constraint
on the annihilation products in $\chi\overline{\chi}$ annihilation.
In general, the positron data would give a weaker constraint as it allows
some level of signals of dark matter annihilation \cite{ours}.

\begin{figure}[t!]
\centering 
\includegraphics[width=7.5in,clip]{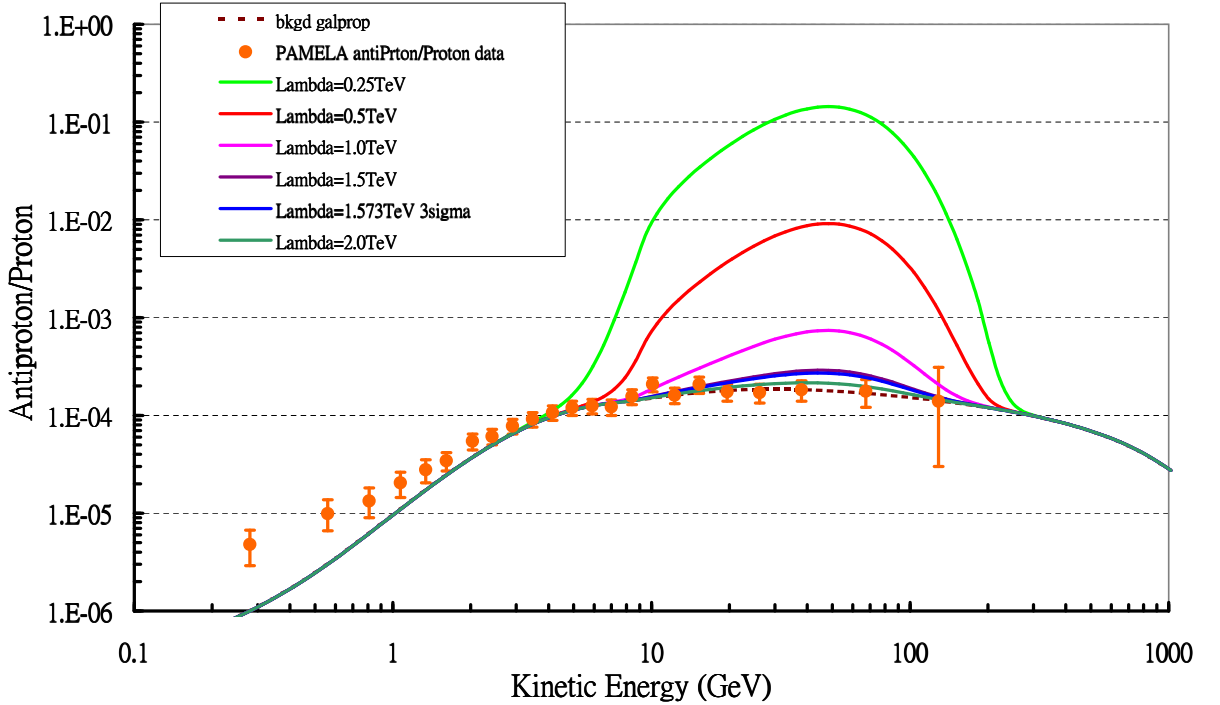}
\caption{\small \label{pbar1}
Antiproton fraction spectrum predicted for the operator 
$O_1 = \frac{1}{\Lambda^2} (\overline{\chi} \gamma^\mu \chi)\, 
 (\bar q \gamma_\mu q)$ for a few values of $\Lambda$. 
The mass of the dark matter is chosen to be 200 GeV here.
The data points are from PAMELA \cite{pamela-p}.
}
\end{figure}

The antiproton flux can be obtained by solving the 
diffusion equation with the corresponding diffusion terms and the 
appropriate source term for the input antiproton spectrum:
\begin{equation}
Q_{\rm ann} = \eta \left( \frac{\rho_{\rm CDM} }{M_{\rm CDM}} \right )^2 
\, \sum \langle \sigma v \rangle_{\bar p} \, \frac{d N_{\bar p}}{ d T_{\bar p} }
 \;,
\end{equation}
where $\eta =1/2\;(1/4) $ for (non-)identical initial state, and $T_{\bar p}$ is 
the kinetic energy of the antiproton
which is conventionally used instead of the total energy. 
We again solve the diffusion equation using GALPROP\,\cite{GALPROP}.

In our case, the dominant contribution comes from
\begin{equation}
\chi \overline{\chi} \to q \bar q \to \bar p + X \;,
\end{equation}
in which all the $q, \bar q\; (q=u,d,c,s,b)$ have probabilities 
fragmenting into $\bar p$.  We adopt a publicly available
code\,\cite{kniehl} to calculate the fragmentation function $D_{q\to h}(z)$ 
for any quark $q$ into hadrons $h$, e.g., $p,\bar p, \pi$.  
The fragmentation
function is then convoluted with energy spectrum $d N/ dT$ 
of the light quark to obtain the energy spectrum of the antiproton
$d N /d T_{\bar p}$.  
The source term $dN /d T_{\bar p}$ is then 
implemented into
GALPROP to calculate the propagation from the halo to the Earth.
We display the energy spectrum for the antiproton fraction 
in Fig.~\ref{pbar1} for the operator $O_1$, in which various values of
$\Lambda$ are chosen.  We only chose a typical operator.  The effects
of other operators are similar.  

Here we adopt a simple statistical measure to quantify the effect
of each operator.  We calculate the $3\sigma$ limit on each scale
$\Lambda_i$. We assume the data agree well with the expected  background,
and then we calculate the $\chi^2$ with finite $\Lambda_i$'s
until we obtain a $\chi^2$ difference of $\chi^2 - \chi^2_{\rm bkgd} = 9$ 
($3\sigma$).  
Note that the uncertainties in the background estimation of the
low energy range ($\alt 4$ GeV) are large in GALPROP, mainly because
of different profiles employed.  We therefore focus on the data points above
4 GeV when we calculate the $\chi^2$.  The data points above 4 GeV
enjoy a small $\chi^2 = 5.0$ for 13 degrees of freedom.
We tabulate all the lower limits of $\Lambda_i$s in Table~\ref{limit}
for $m_\chi$ = 50, 100, 200 and 400 GeV. 

Limits for intermediate values of $m_\chi$ are shown in 
Fig.~\ref{figA}, Fig.~\ref{figB}, and Fig.~\ref{figC} for 
Dirac DM with (axial) vector-boson/tensor-like exchanges, 
Dirac DM with (pseudo) scalar-boson
exchanges and Dirac DM with gluonic interactions, and scalar DM,
respectively.  
The solid lines are the {\it upper} limits due to thermal relic density.
The dashed lines are the {\it lower} limits due to antiproton flux.  
Therefore, for each operator there is a valid range of $\Lambda$.
For example, 
the operator $O_{1,3}$ requires $ 1.6 \; {\rm TeV} \lesssim 
\Lambda_{1,3}
\lesssim 3 \; {\rm TeV}$ for $m_\chi = 200$ GeV. 
The best limit is obtained for the Dirac DM with tensor interactions 
in $O_{5,6}$ at 
$ 1.9\;{\rm TeV} \lesssim \Lambda_{5,6} \lesssim 3.6\;{\rm TeV}$ for 
$m_\chi = 200$ GeV.
In general, the Dirac DM with vector-like exchanges
gives the best limit, except for the operators $O_{2,4}$, which are
well known for velocity suppression. The operators $O_{7-10}$ for Dirac 
DM with scalar-boson exchanges naturally give rather weak limits,
because of the factor $m_q$ in the coupling constant. In addition,
operators $O_{7,9}$ are  suppressed by the velocity. The gluonic
interactions in the operators $O_{11-14}$ also give mild limits because of
the $\alpha_s\approx 10^{-1}$ in the coupling constant, in which
the operators $O_{11,13}$ are further suppressed by velocity. 
On the other hand, the operators for scalar DM give rather mild limits,
especially $O_{15,16}$ give the weakest limits, because the 
derivative couplings in $O_{15,16}$ bring down a factor of momentum.

\begin{table}[th!]
\caption{\small \label{limit}
The $3\sigma$ lower limits on the operators listed in Table~\ref{table1}.
We take the coefficient $C=1$ with $m_\chi$ = 50, 100, 200 and 400 GeV.
We have used the PAMELA data points above the kinetic energy 
$T=4$ GeV in our analysis, because of the large uncertainty of
the theoretical background at low energy. The $\chi^2 ({\rm bkdg})=5.0$.}
\medskip
\begin{tabular}{lrrrr}
\hline
Operators       &   \multicolumn{4}{c}{$\Lambda$ (TeV)} \\
                & $m_\chi$  (GeV)  = 50  & 100   &  200   & 400 \\
\hline
\hline
\multicolumn{4}{c}{Dirac DM, Vector Boson Exchange} \\
\hline
$O_1 = (\overline{\chi} \gamma^\mu \chi)\, (\bar q \gamma_\mu q)$ &
      $1.15$ & $1.34$ & $1.57$ & $1.66$  \\
$O_2 = (\overline{\chi} \gamma^\mu \gamma^5\chi)\, (\bar q \gamma_\mu  q)$ &
       $0.033$ & $0.038$ & $0.045$ & $0.047$ \\
$O_3 = (\overline{\chi} \gamma^\mu \chi)\, (\bar q \gamma_\mu \gamma^5 q)$ &
     $1.15$  &  $1.34$ & $1.57$ & $1.66$ \\
$O_4 = (\overline{\chi} \gamma^\mu \gamma^5 \chi)\, 
    (\bar q \gamma_\mu \gamma^5 q)$ &  
 $0.19$ & $0.15$ & $0.11$ & $0.09$  \\
$O_5 = (\overline{\chi} \sigma^{\mu\nu} \chi)\, (\bar q \sigma_{\mu\nu} q)$ & 
       $1.37$  &  $1.60$  &  $ 1.87 $  & $1.97$ \\
$O_6 = (\overline{\chi} \sigma^{\mu\nu} \gamma^5 \chi)\, 
  (\bar q \sigma_{\mu\nu} q)$ & $1.36$ & $1.60$ &  $1.87 $ & $1.97$\\
\hline
\multicolumn{4}{c}{Dirac DM, Scalar Boson Exchange} \\
\hline
$O_7 = (\overline{\chi}  \chi)\, (\bar q  q)$ & $0.012$ 
                        & $0.013$ & $ 0.014$ & $0.015$\\
$O_8 = (\overline{\chi} \gamma^5  \chi)\, (\bar q  q)$ & 
             $0.12$&    $0.13$ &  $0.14$ & $0.15$ \\
$O_9 = (\overline{\chi}  \chi)\, (\bar q \gamma^5 q)$&
    $0.012$ & $0.013$ &    $ 0.014$ & $0.015$ \\
$O_{10} = (\overline{\chi} \gamma^5 \chi)\, (\bar q \gamma^5 q) $ &  
   $0.12$ &  $0.13$  & $0.14$ & $0.15$ \\
\hline
\multicolumn{4}{c}{Dirac DM, Gluonic} \\
\hline
$O_{11} = (\overline{\chi} \chi)\, G_{\mu\nu} G^{\mu\nu}$ & 
     $0.013$ &  $0.015$  &  $ 0.019$  & $0.027$ \\
$O_{12} = (\overline{\chi} \gamma^5 \chi)\, G_{\mu\nu} G^{\mu\nu} $ & 
     $0.13$ &   $0.15$   &     $ 0.19$  & $0.27$ \\
$O_{13} = (\overline{\chi} \chi)\, G_{\mu\nu} \tilde{G}^{\mu\nu}$ & 
     $0.013$ &   $0.015$  &  $ 0.019 $ & $0.027$ \\
$O_{14} = (\overline{\chi} \gamma^5 \chi)\, G_{\mu\nu} \tilde{G}^{\mu\nu} $& 
     $0.13$ &  $0.15$  &  $ 0.19  $ & $0.27$\\
\hline
\hline
\multicolumn{4}{c}{Complex Scalar DM, Vector Boson Exchange} \\
\hline
$O_{15} = (\chi^\dagger \overleftrightarrow{\partial_\mu} \chi)\, 
  ( \bar q \gamma^\mu q ) $&
    $0.033$ &    $0.038$ &  $0.045$ & $0.047$ \\
$O_{16} = (\chi^\dagger \overleftrightarrow{\partial_\mu} \chi)\, 
( \bar q \gamma^\mu \gamma^5 q ) $&
    $0.033$ &  $0.038$ & $ 0.045  $ & $0.047$ \\
\hline
\multicolumn{2}{c}{\;\;\;\;\;\;\;\;\;Complex Scalar DM, Scalar Vector Boson Exchange} \\
\hline
$O_{17} = (\chi^\dagger \chi)\, ( \bar q  q ) $&
    $0.16$ &     $0.13$   &        $ 0.099 $ & $0.074$ \\
$O_{18} = (\chi^\dagger \chi)\, ( \bar q  \gamma^5 q ) $&
    $0.16$ &     $0.13$     &  $ 0.099   $ & $0.074$ \\
\hline
\multicolumn{4}{c}{Complex Scalar DM, Gluonic} \\
\hline
$O_{19} = (\chi^\dagger \chi)\, G_{\mu\nu} {G}^{\mu\nu} $   & 
     $0.18$ &       $0.15$ &  $ 0.15  $ & $0.18$ \\
$O_{20} = (\chi^\dagger \chi)\, G_{\mu\nu} \tilde{G}^{\mu\nu} $& 
     $0.18$ &      $0.15$      &     $  0.15  $& $0.18$ \\
\hline
\hline
\end{tabular}
\end{table}

\section{Discussion and Conclusions}

Here we do a comparison with the limits obtained in Ref.~\cite{tait-gamma},
in which limits from relic density, Tevatron, and gamma-ray are shown.
Comparisons are summarized as follows.
\begin{enumerate}

\item
The limits due to relic density obtained in this work are consistent
with results of Ref.~\cite{tait-gamma}.

\item 
In Fig.~5 of  Ref.~\cite{tait-gamma}, the limits for their $D_{1-4}$ 
(corresponding to our $O_{7-10}$,
 Dirac DM with (pseudo) scalar-boson exchanges) are shown.  
The limits obtained from FERMI gamma-ray are about the same as
what we obtained from PAMELA antiproton data.  The limits from FERMI 
improve with increasing $m_\chi$ while it is almost flat in our case 
(see Fig.~\ref{figB}).  

\item
In Fig.~6 of  Ref.~\cite{tait-gamma}, the limits for their $D_{5-8}$
(corresponding to our $O_{1-4}$, Dirac DM with (axial) vector-boson exchanges) 
are shown.  The limits on $O_{1,3}$ obtained from FERMI
gamma-ray are 
about $0.1 - 0.5$ TeV while the limits that we obtained from 
antiproton data are $1.1 - 1.7$ TeV, significantly stronger.  
The limits on $O_{2,4}$ are velocity suppressed and are 
$0.03 - 0.2$ TeV.

\item 
In Fig.~8 of  Ref.~\cite{tait-gamma}, the limits for their $C_{1,2}$
(corresponding to our $O_{17,18}$, complex scalar DM with scalar-boson 
exchanges) are shown.  The limits that we obtained from antiproton data
are slightly stronger than those from FERMI gamma-ray data.

\item
In Fig.~9 of  Ref.~\cite{tait-gamma}, the limits for their $C_{3,4}$
(corresponding to our $O_{15,16}$, complex scalar DM with vector-boson 
exchanges) are shown.  The limits that we obtained from antiproton data
are weaker than those from FERMI gamma-ray data.

\item The limits from Tevatron data \cite{tait-gamma} are rather insensitive
to $m_\chi$, except when $m_\chi \agt 200$ GeV.  The limits obtained from
FERMI gamma-ray
data roughly improve with increasing $m_\chi$ \cite{tait-gamma}.
The limits obtained from PAMELA antiproton data are in general quite flat.

\end{enumerate}

In summary, we have used an effective interaction approach to investigate
the effects of dark matter interactions with light quarks on antiproton flux
from the Galactic halo.  We have assumed a standard halo density 
and used
the GALPROP to calculate the diffusion.  The obtained antiproton flux is
then compared with the PAMELA data.  We have successfully used the data to
obtain a $3\sigma$ limits on the scale $\Lambda_i$. The best limits
are from the Dirac DM with vector-boson or tensor-like exchanges.  The
limits are about $1 - 2$ TeV.  While the other operators give milder limits.
Note that these limits from antiproton flux are {\it lower} limits on
$\Lambda_i$.
With the requirement of not exceeding the relic density of the
cold dark matter deduced from the WMAP, we also obtain the {\it upper} 
limits on $\Lambda_i$.
Therefore, both the relic density and antiproton constraints give
a valid range for each 
$\Lambda_i$, e.g, $ 1.6 \; {\rm TeV} \lesssim \Lambda_1
\lesssim 3 \; {\rm TeV}$ for $m_\chi = 200$ GeV. This is a very useful piece of 
information on the effective interactions of dark matter with the
SM light quarks that can give useful implications for
collider searches and direct detection.

\nopagebreak

\appendix
\section{ Annihilation Cross Sections}
Here we list all the formulas for annihilation cross sections
of the operators $O_1$ to $O_{20}$.  
\begin{eqnarray}
\frac{d\sigma_1}{dz} &=& \frac{1}{\Lambda^4} \frac{N_C}{16 \pi s}
\frac{\beta_q}{\beta_\chi} \left[ u_m^2 + t_m^2 + 2s (m_\chi^2 +m_q^2)\right ]
                     \;, \\
\frac{d\sigma_2}{dz} &=& \frac{1}{\Lambda^4} \frac{N_C}{16 \pi s}
   \frac{\beta_q}{\beta_\chi} \left[ u_m^2 + t_m^2 + 2s (m_q^2 - m_\chi^2)
          - 8 m_q^2 m_\chi^2 \right ] \;,  \\
\frac{d\sigma_3}{dz} &=& \frac{1}{\Lambda^4} \frac{N_C}{16 \pi s}
   \frac{\beta_q}{\beta_\chi} \left[ u_m^2 + t_m^2 + 2s (m_\chi^2 - m_q^2)
          - 8 m_q^2 m_\chi^2 \right ] \;,  \\ 
\frac{d\sigma_4}{dz} &=& \frac{1}{\Lambda^4} \frac{N_C}{16 \pi s}
   \frac{\beta_q}{\beta_\chi} \left[ u_m^2 + t_m^2 - 2s (m_\chi^2 + m_q^2)
          +16 m_q^2 m_\chi^2 \right ] \;,  \\  
\frac{d\sigma_5}{dz} &=&  \frac{1}{\Lambda^4} \frac{N_C}{4 \pi s}
  \frac{\beta_q}{\beta_\chi} \left[2(u_m^2 + t_m^2) + 2s (m_\chi^2 +m_q^2) 
   + 8 m_q^2 m_\chi^2 - s^2\right ] \;, \\
\frac{d\sigma_6}{dz} &=&  \frac{1}{\Lambda^4} \frac{N_C}{4 \pi s}
  \frac{\beta_q}{\beta_\chi} \left[2(u_m^2 + t_m^2) + 2s (m_\chi^2 +m_q^2) 
   -16 m_q^2 m_\chi^2 - s^2\right ] \;, \\
\frac{d\sigma_7}{dz} &=&  \frac{m_q^2}{\Lambda^6} \frac{N_C}{32 \pi} 
  s \beta_\chi \beta_q^3 \;, \\
\frac{d\sigma_8}{dz} &=&   \frac{m_q^2}{\Lambda^6} \frac{N_C}{32 \pi} 
   \frac{s \beta_q^3 }{\beta_\chi} \;, \\
\frac{d\sigma_9}{dz} &=&  \frac{m_q^2}{\Lambda^6} \frac{N_C}{32 \pi} 
  s \beta_\chi \beta_q \;, \\
\frac{d\sigma_{10}}{dz} &=& \frac{m_q^2}{\Lambda^6} \frac{N_C}{32 \pi} 
   \frac{s \beta_q }{\beta_\chi} \;, \\  
\frac{d\sigma_{11}}{dz} &=&   \frac{\alpha_s^2}{\Lambda^6} \frac{1}{32 \pi} 
  s^2 \beta_\chi \;, \\
\frac{d\sigma_{12}}{dz} &=&   \frac{\alpha_s^2}{\Lambda^6} \frac{1}{32 \pi} 
  \frac{s^2}{\beta_\chi} \;, \\
\frac{d\sigma_{13}}{dz} &=&  \frac{d\sigma_{11}}{dz} \;,\\
\frac{d\sigma_{14}}{dz} &=&  \frac{d\sigma_{12}}{dz} \;,\\
\frac{d\sigma_{15}}{dz} &=&  \frac{1}{\Lambda^4}\frac{N_C}{4\pi s}
   \frac{\beta_q}{\beta_\chi} ( u t - m_q^2(u+t) - m_\chi^4 +m_q^4 ) \;, \\
\frac{d\sigma_{16}}{dz} &=&  \frac{1}{\Lambda^4}\frac{N_C}{4\pi s}
   \frac{\beta_q}{\beta_\chi} \left( u t - (m_\chi^2 - m_q^2 )^2 \right) \;, \\
\frac{d\sigma_{17}}{dz} &=&  \frac{m_q^2}{\Lambda^4}\frac{N_C}{16\pi}
   \frac{\beta_q^3}{\beta_\chi} \;, \\
\frac{d\sigma_{18}}{dz} &=&  \frac{m_q^2}{\Lambda^4}\frac{N_C}{16\pi}
   \frac{\beta_q}{\beta_\chi} \;, \\ 
\frac{d\sigma_{19}}{dz} &=&   \frac{\alpha_s^2}{\Lambda^4} \frac{1}{16 \pi} 
  \frac{s}{\beta_\chi} \;, \\
\frac{d\sigma_{20}}{dz} &=&  \frac{d\sigma_{19}}{dz} \;,
\end{eqnarray}
where $s$ is the square of 
the center-of-mass energy, $z$ is the cosine of scattering angle,
$u_m = u - m_\chi^2 -m_q^2$, $t_m = t - m_\chi^2 - m_q^2$, 
$\beta_\chi = (1 - 4 m_\chi^2/s)^{1/2}$, $\beta_q = (1-4 m_q^2/s)^{1/2}$,
and $N_C$ is the color factor (3 for quarks). We have set
the coefficient $C=1$ in these formulas. 
The annihilation rate in the nonrelativistic limit will then be given by
$\sigma \cdot (2\beta_\chi)$.

\section*{Acknowledgments}
The work was supported in parts by the National Science Council of
Taiwan under Grant Nos. 99-2112-M-007-005-MY3, and
98-2112-M-001-014-MY3, the NCTS, and the
WCU program through the KOSEF funded by the MEST (R31-2008-000-10057-0).

\end{document}